\documentclass[aps,twocolumn,floatfix, showkeys, superscriptaddress, nofootinbib]{revtex4}
\usepackage{amsmath}
\usepackage{graphicx,bm,hhline}
\usepackage{multirow}
\DeclareMathAlphabet{\pazocal}{OMS}{zplm}{m}{n}

\newcommand{\wig}[1]{\mathrel{\hbox{\hbox to 0pt{\lower.6ex\hbox{$\sim$}\hss}\raise.4ex\hbox{$#1$}}}}

\makeatletter
\renewcommand*\env@matrix[1][\arraystretch]{%
  \edef\arraystretch{#1}%
  \hskip -\arraycolsep
  \let\@ifnextchar\new@ifnextchar
  \array{*\c@MaxMatrixCols c}}
\makeatother

\begin{document}
\title{Tabular Electrical Conductivity for Aluminum}

\author{C. E. Starrett}
\email{starrett@lanl.gov}
\affiliation{Los Alamos National Laboratory, P.O. Box 1663, Los Alamos, NM 87545, U.S.A.}

\author{R. Perriot}
\affiliation{Los Alamos National Laboratory, P.O. Box 1663, Los Alamos, NM 87545, U.S.A.}

\author{N. R. Shaffer}
\affiliation{Los Alamos National Laboratory, P.O. Box 1663, Los Alamos, NM 87545, U.S.A.}

\author{T. Nelson}
\affiliation{Los Alamos National Laboratory, P.O. Box 1663, Los Alamos, NM 87545, U.S.A.}

\author{L. A. Collins}
\affiliation{Los Alamos National Laboratory, P.O. Box 1663, Los Alamos, NM 87545, U.S.A.}

\author{C. Ticknor}
\affiliation{Los Alamos National Laboratory, P.O. Box 1663, Los Alamos, NM 87545, U.S.A.}


\date{\today}
\begin{abstract}
A new Sesame-type table for the electrical conductivity of aluminum is described.  The table
is based on density functional theory calculations and ranges from 10$^{-3}$ to 1 times
solid density (2.7 g/cm$^3$), and from 10$^{-2}$ to 10$^3$ eV in temperature.  The table is compared
to other simulations and to experiments and is generally in good agreement.  The high temperature,
classical limit of the conductivity is recovered for the highest temperatures and lowest densities.
The table is critically evaluated and directions for improvements are discussed.
\end{abstract}
\pacs{ }
\keywords{electrical conductivity}
\maketitle

\section{Introduction}
Electrical conductivity in dense materials is an important physical parameter for a variety
of applications.  For example, it is needed for modelling magnetic liner fusion experiments \cite{gomez14} and
pulse power experiments \cite{stygar15}, or the dynamo effect in the core of planets \cite{stacey01}.

In the low density, high temperature regime, classical methods using Coulomb Logarithms 
are accurate \cite{cohen50, lee84, braginskii58} and quick to evaluate.  However, in the dense regime these approaches
break down due to effects like ion correlation, partial ionization, core-valence orthogonality, degeneracy,
and multiple scattering effects \cite{starrett18a}.  

An alternative approach is to use accurate models, that are too expensive and complicated to be used directly in
applications, to make data tables.  For example, Rinker \cite{rinker88} made Sesame \cite{lyon92} data tables based on
an average atom approach coupled to the Ziman approximation \cite{ziman60}.  This method is reasonable
in the degenerate regime (high density, low temperature), but is inaccurate elsewhere.  Desjarlais \cite{desjarlais01}
also made Sesame tables by adjusting the Coulomb Log based Lee-More model \cite{lee84} to better match
available experimental data among other improvements.  Another approach is to use a chemical-model \cite{kuhlbrodt05}, which can be
rapid to evaluate but does not include some relevant physics like ionic structure (for a review see reference \cite{redmer97}).

We take the approach of using high fidelity density functional theory (DFT) based calculations to model
the entire density-temperature plane of interest where the material is a fluid.  We focus on aluminum,
as it is a widely used conductor, but the approach can be applied to other materials.  We employ two methods:
the accurate but expensive density functional theory molecular dynamics (DFT-MD), coupled to the Kubo-Greenwood approximation \cite{desjarlais02}, which
is practical for degenerate systems, and
the more approximate Potential of Mean Force approach \cite{starrett17}, which is accurate for moderate
to weakly degenerate plasmas and is relatively inexpensive.

It is also possible to use the Kubo-Greenwood method with the average atom model
\cite{starrett12a, gill19, faussurier14, ovechkin16, sterne07, johnson}, but we have opted not to use this approach as it remains
to be proven that this contains the correct limiting behaviour at high temperature, and is more computationally expensive.  In contrast, the 
potential of mean force approach which relies on the relaxation time approximation does recover the classical
limit \cite{starrett18} and is rapid to evaluate.

\section{Construction of table}
The table ranges from 10$^{-3}$ to 1 times solid density, taken here as 2.7 g/cm$^3$, and from 10$^{-2}$ to
10$^3$ eV in temperature.  We start by calculating the electrical conductivity for the full range of the table
using the potential of mean force model of reference \cite{starrett17}.  This model uses the BGK \cite{bhatnager54}
approximation (also know as the relaxation time approximation).  The electron relaxation time is
calculated using the quantum mechanical expression for the momentum transport cross
section
\begin{equation}
\sigma_{\mathrm{tr}}(\epsilon) = \frac{4 \pi \hbar^2}{p^2} \sum\limits_{l=0}^{\infty} (l+1) \left( \sin\left( \eta_{l+1} - \eta_{l} \right)\right)^2
\label{str}
\end{equation}
where $p$ is the electron momentum, the $\eta_l$ are energy dependent phase shifts and the sum 
over angular momentum quantum number $l$ converges.  The phase shifts are evaluated by solving the
Schr\"odinger equation for the scattering potential.  This potential, the potential of mean force $V^{MF}(r)$,
is calculated using the finite temperature DFT based average atom-two component
plasma model (AA-TCP) developed in references \cite{starrett13,starrett14}.  The potential of mean
force recovers the Debye-Huckel potential at high temperature and low density.  It takes into
account core states, core-valence orthogonality, ionic structure (through the pair distribution
function), and partial ionization.

This approach to the conductivity becomes inaccurate in systems where the density of states of
scattering electrons deviates strongly from free-electron like behaviour.  For example, in the expanded
metal regime, with temperatures of $\sim$ 1 eV, and densities $\sim$ 1/10$^{th}$ of solid (i.e. warm 
dense matter), where electron states are relocalizing as density is reduced, this model will be
inaccurate (see \cite{starrett17}).
In general, however, this model is reasonably accurate and computationally inexpensive, allowing construction
of wide ranging tables.  Since this is a DFT-based model we must choose an exchange and correlation potential.
We have used the temperature dependent LDA parameterization of reference \cite{ksdt}.
For the electron-electron contribution to the electrical conductivity we have used the formula
given in reference \cite{reinholz15}.
\begin{figure}
\begin{center}
\includegraphics[scale=0.3]{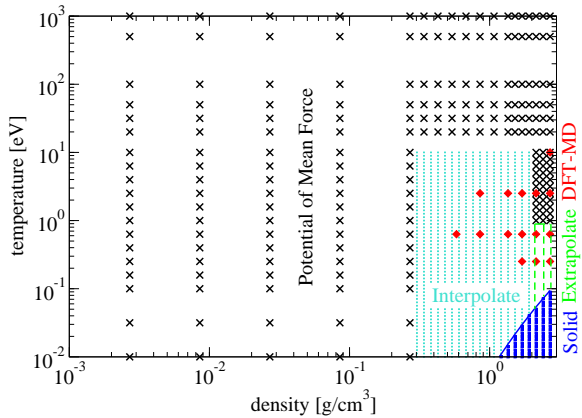}
\end{center}
\caption{(Color online) 
Graphic describing the construction of the table.
Black $\times$'s are density-temperature ($\rho-T$) points for which the potential of mean force model $V^{MF}$
\cite{starrett17} was evaluated.  Red diamonds are $\rho-T$ points where we evaluated DFT-MD.
The dark blue region in the bottom right hand corner is where the solid model is used.  The light blue
and green regions correspond to regions of interpolation \cite{akima70} and extrapolation which were
guided by the DFT-MD and $V^{MF}$ results.
}
\label{fig_models}
\end{figure}

\begin{figure}
\begin{center}
\includegraphics[scale=0.4]{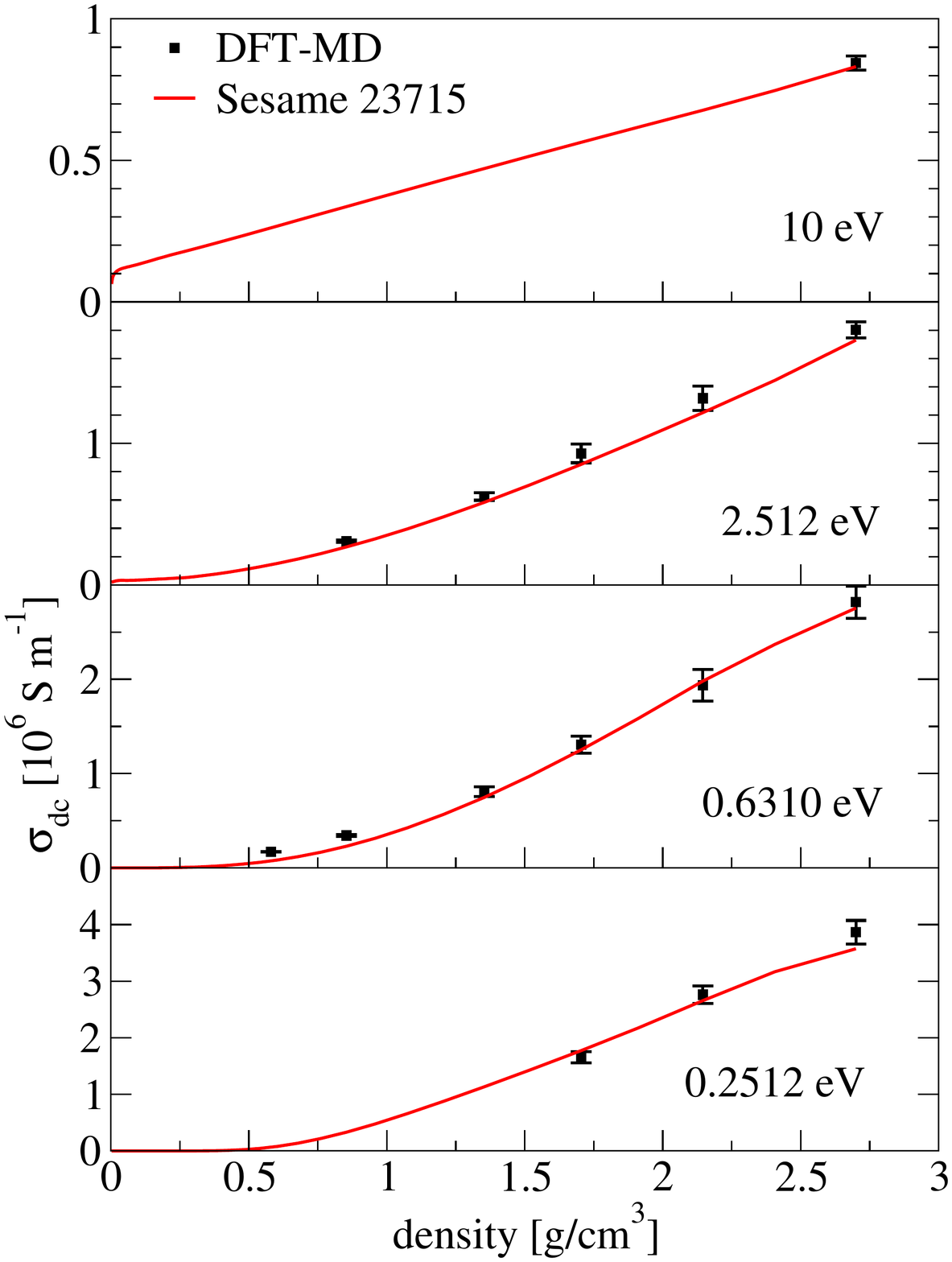}
\end{center}
\caption{(Color online) Comparison of new table to DFT-MD results using PBE \cite{pbe}.
}
\label{fig_it}
\end{figure}
\begin{figure}
\begin{center}
\includegraphics[scale=0.3]{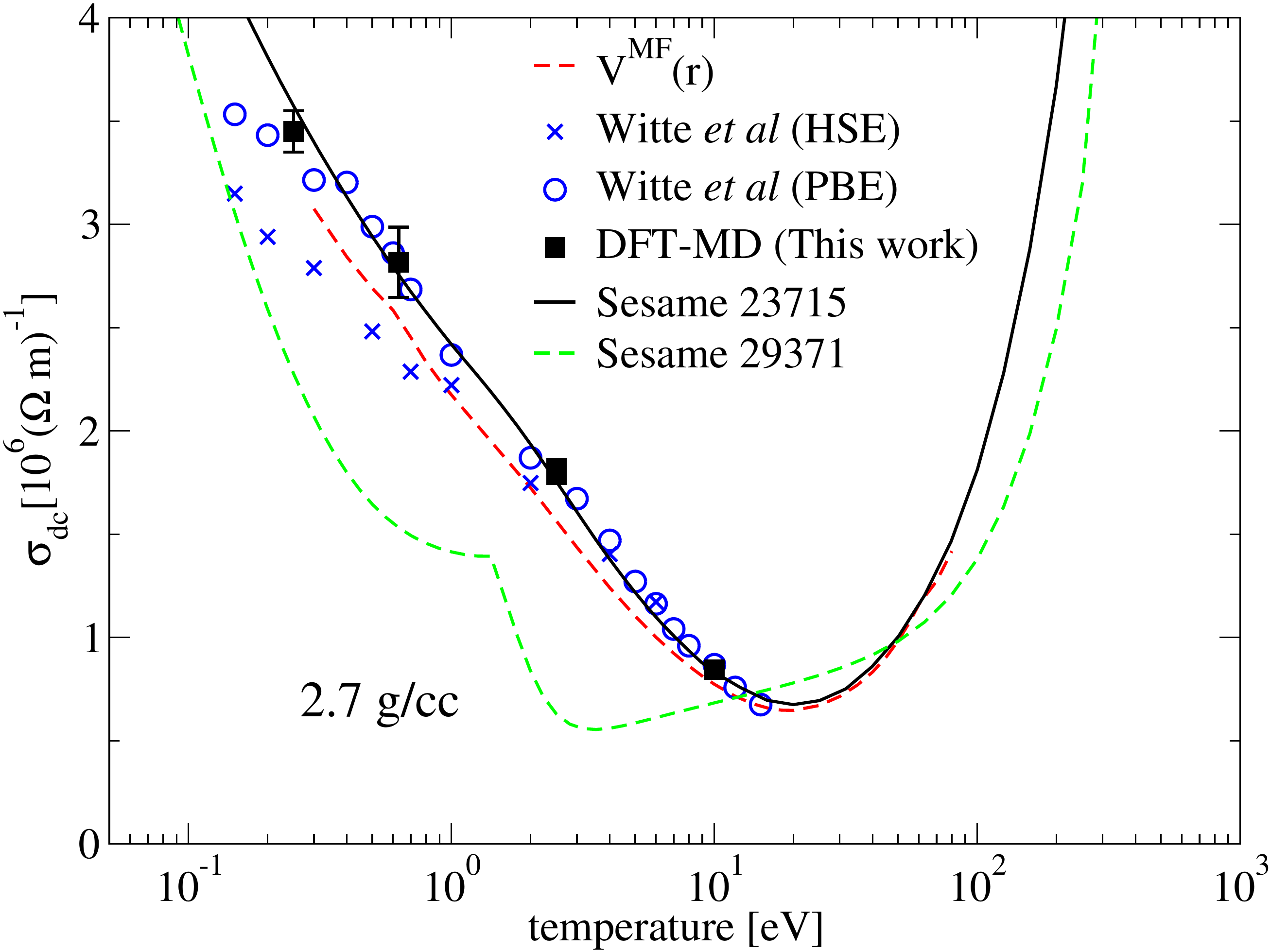}
\end{center}
\caption{(Color online)  Isochore (2.7 g/cm$^3$) comparing the new table DFT-MD (this work),
Sesame 29371 \cite{desjarlais01}, DFT-MD due to Witte et al \cite{witte18}, and the model
of reference \cite{starrett17} ($V^{MF}(r)$).
}
\label{fig_witte}
\end{figure}
\begin{figure}
\begin{center}
\includegraphics[scale=0.3]{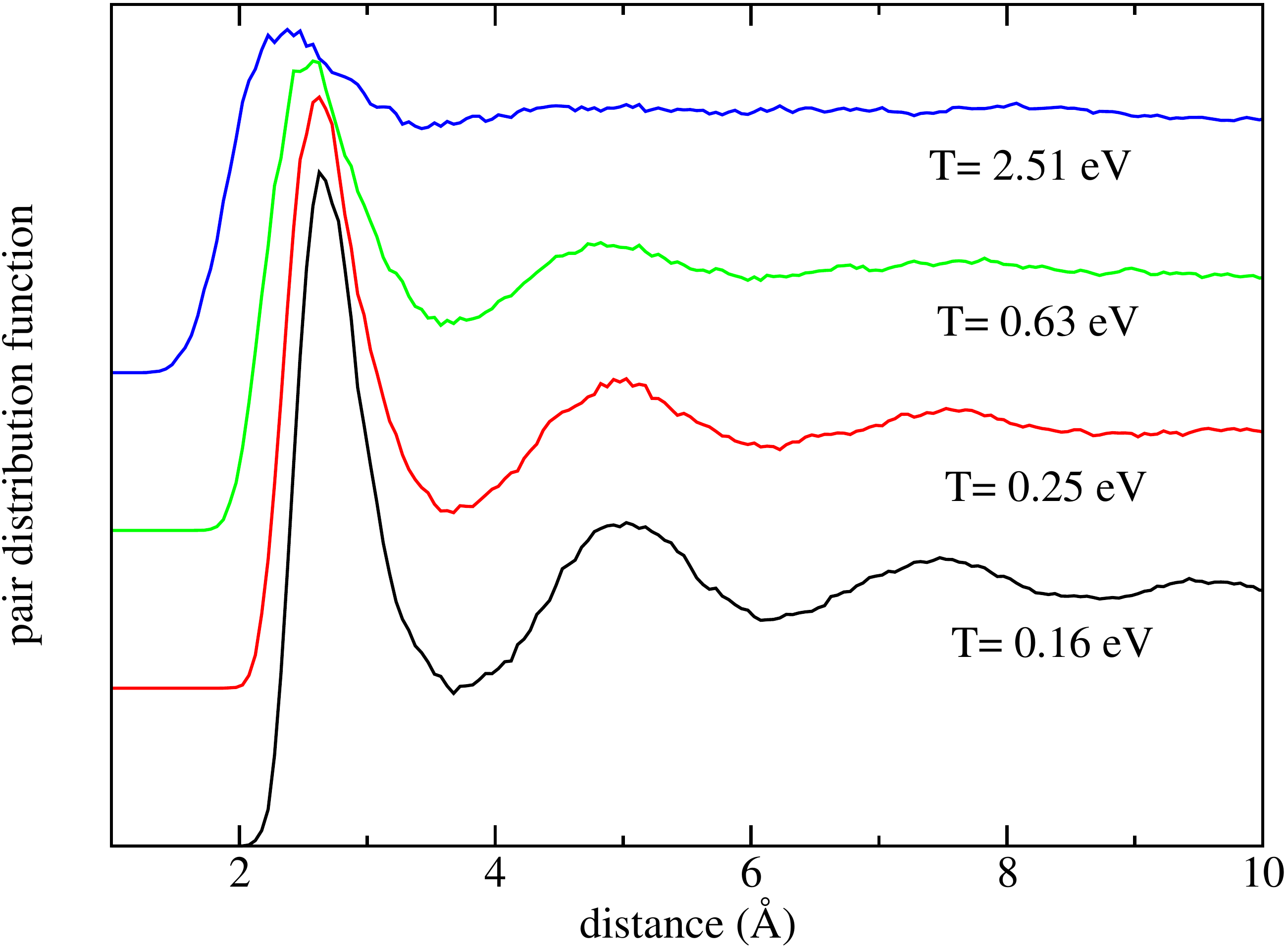}
\end{center}
\caption{(Color online)  DFT-MD pair distribution functions along an isochore (2.7 g/cm$^3$).  
The curves are shifted arbitrarily along the vertical axis for clarity.  At the lowest temperatures
the pair distribution function displays significant long range structure.
}
\label{fig_qmdgr}
\end{figure}

As a more accurate approach in the warm dense matter regime, we employ  quantum molecular dynamics (QMD) in the Born-Oppenheimer approximation, which decouples the ionic from the electronic components. The ions move according to classical equations of motion governed by forces due to the ions and electrons. The electrons receive a quantum mechanical treatment based on temperature-dependent density functional theory (TD-DFT)\cite{martin04}. We follow the Kohn-Sham (KS) implementation and solve for the electronic orbitals over a set of atoms in a periodically-replicated cell; these orbitals in turn determine the electrical conductivity through the Kubo-Greenwood (KG) approximation \cite{greenwood58}. 

We have performed DFT-MD calculations with the Vienna ab-initio simulation package 
(VASP~\cite{kresse93, kresse94, kresse96, kresse96a}), using the Generalized Gradient Approximation in the Perdew, 
Burke, Ernzerhof formulation for the exchange-correlation (XC) functional (GGA-PBE)~\cite{pbe,pbe2} and a GW three-electron (3e) plane augmented 
wave (PAW) pseudopotential (PP) ~\cite{blochl94,kresse99} with a plane-wave cutoff energy of 750 eV. All of the DFT-MD 
simulations were performed at constant temperature using the Nose-Hoover thermostat with a time step of 1-2 fs at a single k-point (Gamma), containing all orbitals
or bands with Fermi-Dirac occupation greater than 1$\times10^{-4}$.

Samples containing 64 atoms at density $\rho$ were first equilibrated at temperature 
T for at least 2 ps. From this trajectory, we extracted five to ten snapshots separated by a time interval of roughly 2 $\tau$ in order to determine the
 frequency-dependent (ac) electrical conductivity by the KG analysis for temperatures between 0.25eV and 10eV. Snapshots separated by the characteristic or e-folding time $\tau$ for the velocity
autocorrelation function of the self-diffusion coefficient are generally considered  uncorrelated ($\tau \approx$ 20-140 
ps depending on the density and temperature). 
For the optical analysis, we employed the Gamma-point or a 2x2x2 Monkhorst-Pack (MP) grid with four irreducible k-points with the number of bands from twice to three times the MD choice (set I).  
We have also tested convergence using
a 4x4x4 MP grid with 32 k-points, a GW 11e PAW, and a samples of 128 and 250 atoms (set II). We find for temperatures above  0.5eV, the canonical
choice of 64 atoms, GGA-PBE XC functional, GW 3e PAW, and 4 k-points produces results converged within a few percent those of set II for the dc conductivities.
For temperatures below 0.5eV, we employ set II although set I gives conductivities within 10\%.

We use this method as a benchmark and adjust the table to fit these calculations.  Figure
\ref{fig_models} shows the points calculated by these models.  Most of the temperature-density plane
is covered by the potential of mean force model.  In the warm dense matter regime the DFT-MD
results are used to guide the interpolation and extrapolation regimes where the $V^{MF}$ model
is less accurate.  Figure \ref{fig_it} shows the comparison of the new table (Sesame 23715)
to our DFT-MD calculations.  By construction, the agreement is very good.

In figure \ref{fig_witte}, we examine the 2.7 g/cm$^3$ isochore, which displays the new table (Sesame 23715),
our DFT-MD calculations, DFT-MD calculations of Witte et al. \cite{witte18} using two different
exchange-correlation potentials (PBE \cite{perdew96} and HSE \cite{hse}).   
Our PBE results agree with those of Witte et al. \cite{witte18} over the entire temperature range. We also find the need to employ larger k-point  and atom samples to
obtain converged results below about 0.5eV. This requirement could arise from several conditions, for example, the advent of chemical processes that require more subtle representations of the
medium and reactions or the proximity to melt ($\approx$ 1000K).  Simulations around melt are notably 
difficult to perform with methods such as DFT; also, the external constraint applied by the periodic boundary conditions
on a small cell can lead to artificial effects, such that representing a disordered (liquid) system under these 
conditions becomes nontrivial.  For example, we show in Fig. \ref{fig_qmdgr} by means of the radial distribution function 
at $\rho$=2.7 g/cm$^3$, the increase in structural features in the form of nearest-neighbor peaks as the temperature decreases.

The potential of mean
force results ($V^{MF}$) are also in reasonable agreement with the DFT-MD (figure \ref{fig_witte}), and the agreement 
improves as temperature increases.  The new table (by design) matches the DFT-MD using PBE at
the lower temperatures and the $V^{MF}$ results at high temperatures.  Also shown is
the older Sesame 29371 based on the model \cite{lee84,desjarlais01}.  Some significant 
differences between the Tables are observed ($>$ 50 \%) that could be important in applications (e.g. \cite{sinars11}).

\begin{figure}
\begin{center}
\includegraphics[scale=0.4]{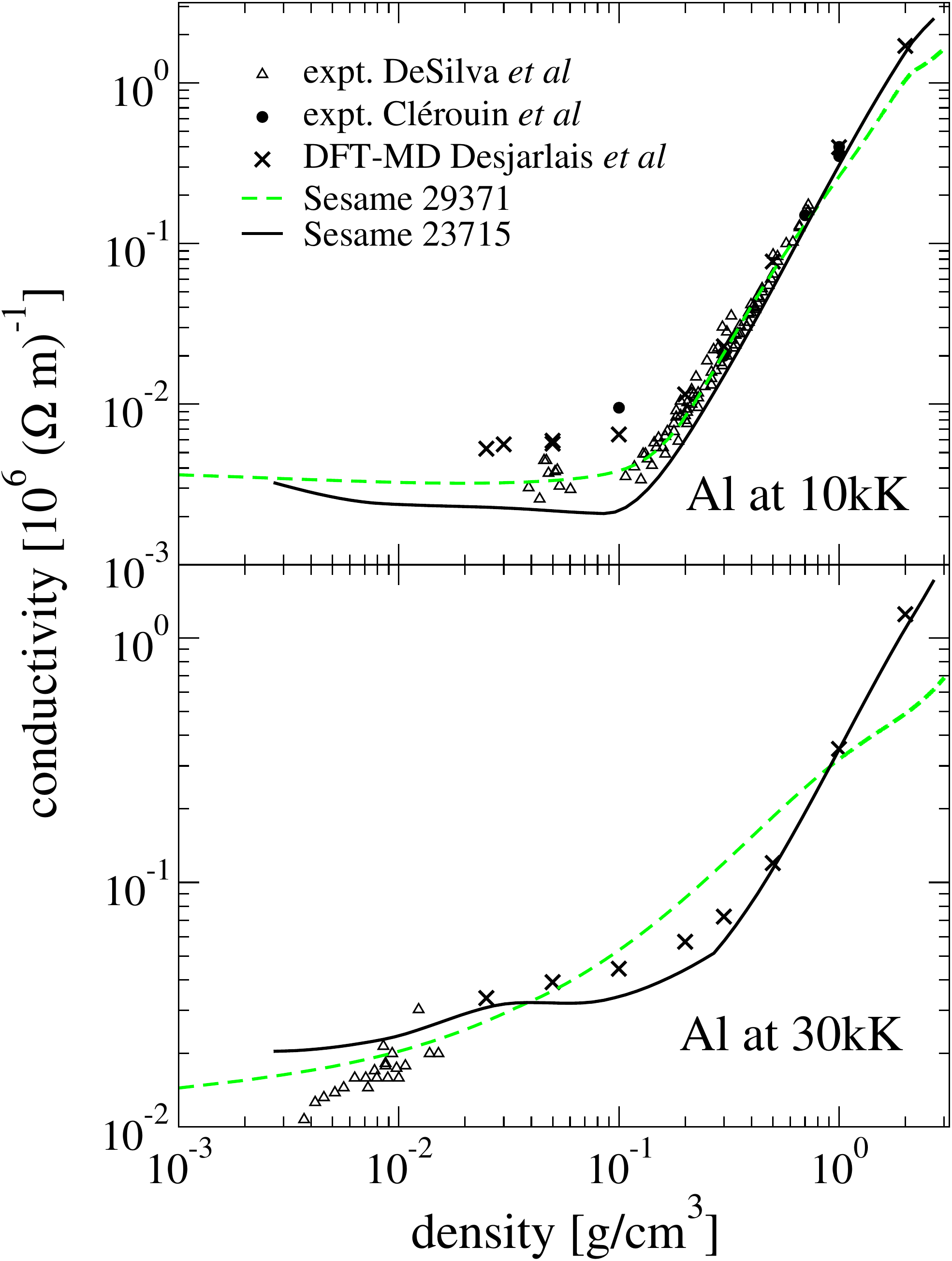}
\end{center}
\caption{(Color online) Isotherms (10 kK $\approx$ 0.86 eV, 30 kK $\approx$ 2.59 eV) for
expanded aluminum compared to the experiments of DeSilva et al \cite{desilva98}, experiment of 
Cl\'erouin et al \cite{clerouin08}, and DFT-MD of Desjarlais \cite{desjarlais02}.  
}
\label{fig_desilva}
\end{figure}
\begin{figure}
\begin{center}
\includegraphics[scale=0.3]{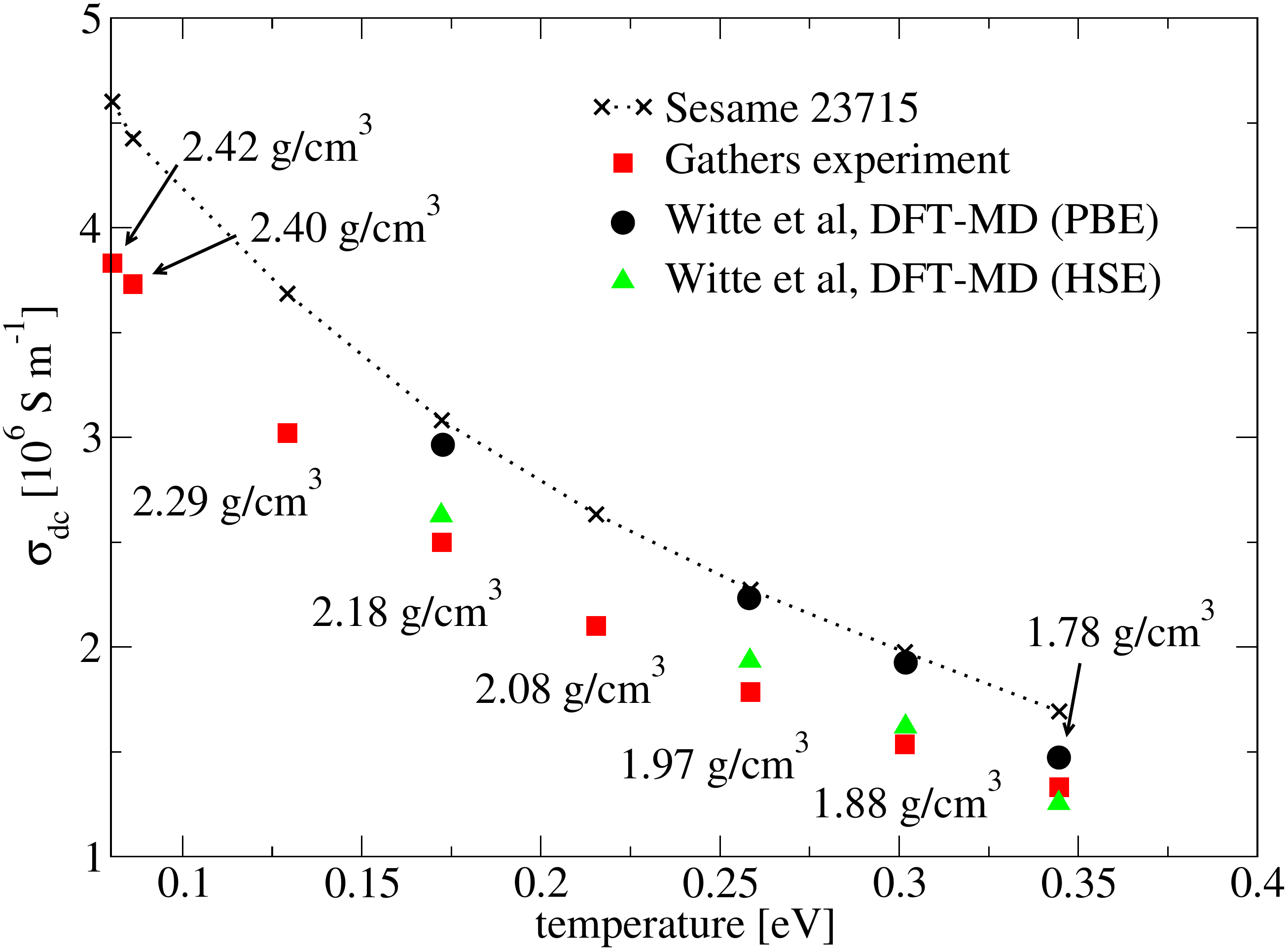}
\end{center}
\caption{(Color online) Comparison to the experiments of Gathers \cite{gathers83}.  These
data have recently been confirmed by another experiment \cite{leitner17}.  Also show are DFT-MD results
of Witte et al \cite{witte18}.   The line connecting the Sesame 23715 points serves only as a
guide to the eye.
}
\label{fig_gathers}
\end{figure}
\begin{figure}
\begin{center}
\includegraphics[scale=0.3]{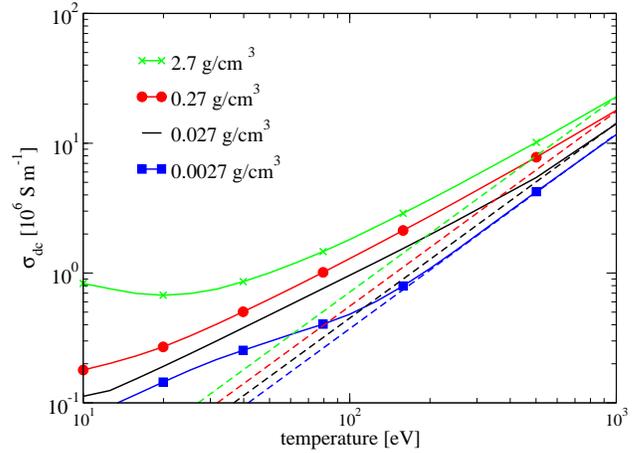}
\end{center}
\caption{(Color online) Asymptotic behavior of the new table.  The dashed lines have
slope $T^{3/2}$ (which is correct in the classical limit), and are fitted to match the 
table at 1000 eV.  At the lowest density and highest temperatures the table recovers this slope.  
}
\label{fig_asym}
\end{figure}

In figure \ref{fig_desilva}, we compare the new table to the experiments of DeSilva et al \cite{desilva98}
and Cl\`erouin et al \cite{clerouin08}.  These experiments test the difficult expanded metal regime
at relatively low temperature, and test our interpolation regime (figure \ref{fig_models}), where
we have used Akima interpolation \cite{akima70}.  The
agreement is reasonable, given that there is a significant discrepancy between the experiments 
themselves, and with the DFT-MD of reference \cite{desjarlais02}.  The results of the $V^{MF}$ model
(not shown, but identical to the new table for densities $<$ 0.27 g/cm$^3$) are sensitive to the
choice of exchange and correlation potential in this regime.  Here we have used the temperature dependent LDA
\cite{ksdt}.  In reference \cite{starrett17} we used a zero temperature parameterization \cite{pz},
and found better agreement with the DFT-MD (which also used a zero temperature parameterization).
The reason for this sensitivity is the low ionization of the system, and hence the small number of
conducting electrons.  A small absolute change in this number caused by changing the exchange and correlation
potential is a significant relative change, hence the effect on the conductivity.  We have chosen
to use the more physically complete temperature dependent exchange and correlation potential \cite{ksdt}.
Also shown in the figure is the older Sesame 29371 table.  It agrees well with the DeSilva et al data
by construction but deviates from the DFT-MD at higher densities.

The new table is compared to experiments in the liquid regime in figure \ref{fig_gathers}.  These
data have recently been confirmed in reference \cite{leitner17}.  Also shown are
DFT-MD results from reference \cite{witte18}.  The new table agrees well with the PBE
calculations, as expected since it was constructed to match our own PBE calculations
in this regime (but not these exact points).  The table and the PBE results
overestimate the data.  HSE calculations by the same authors \cite{witte18} demonstrate
sensitivity to the exchange and correlation potential in this regime.

At high temperatures and low densities the conductivity should be well described by the
Spitzer model \cite{cohen50} where the conductivity increases proportionally
to $T^{\frac{3}{2}}$ along an isochore.  In figure \ref{fig_asym} isochores from the table are shown.  At
the lowest density ($10^{-3}$ times solid density), the model clearly reaches this classical
result.  This behaviour is not enforced, but rather is a consequence of the BGK model
coupled to the quantum cross section (equation (\ref{str})) going over to its classical limit.  From
the figure, as density is increased, this classical behaviour is reached at increasingly
high temperature, and for the two highest densities the classical behaviour is not reached
by 1000 eV.


\begin{figure}
\begin{center}
\includegraphics[scale=0.3]{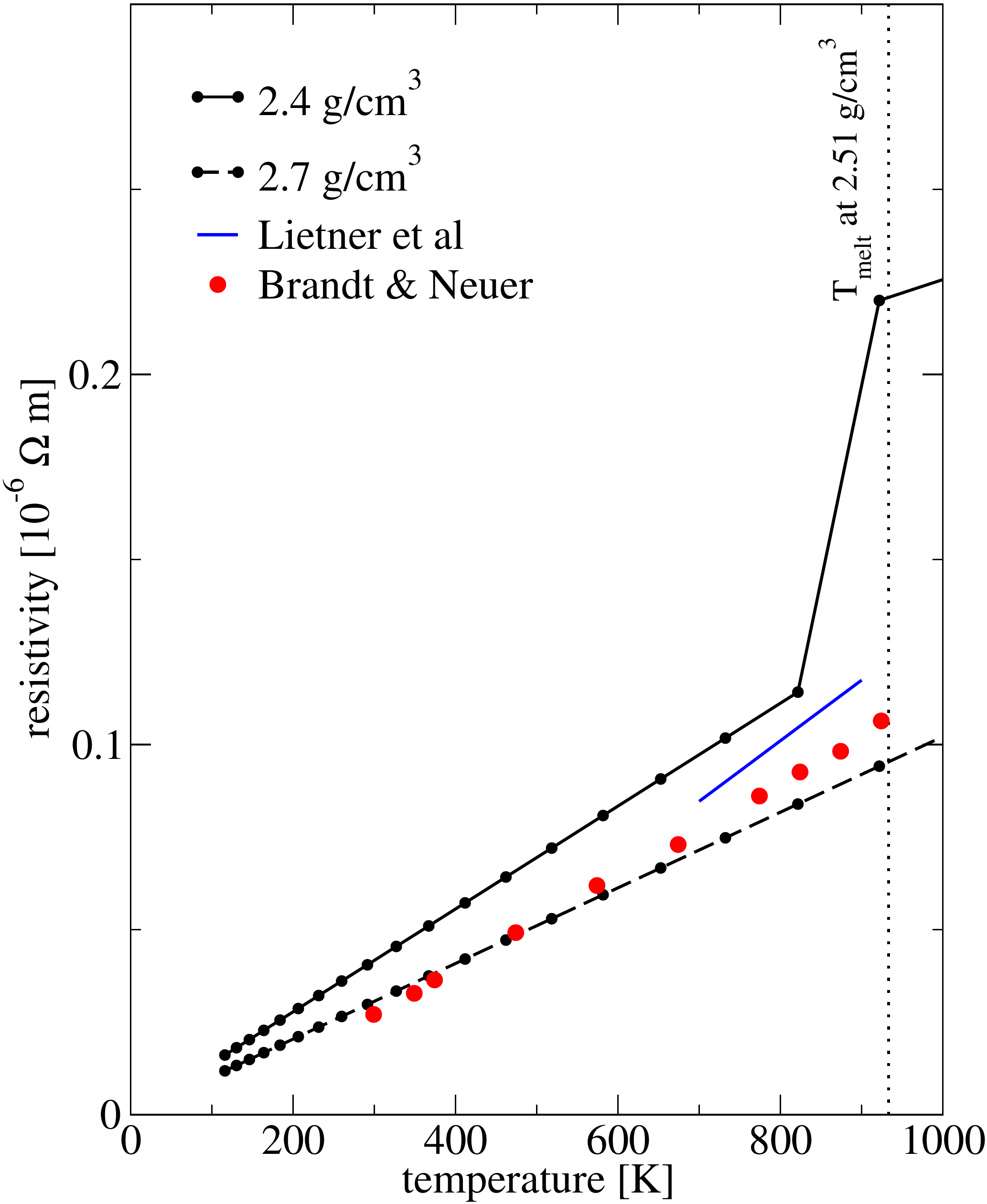}
\end{center}
\caption{(Color online) Solid aluminum resistivity.  Shown are experimental points due to \cite{brandt07}
and a recent set of measurements due to Leitner et al \cite{leitner17},
where we have shown the fit given in their paper and the density runs from 2.53 to 2.63 g/cm$^3$ for the line shown.
Two adjacent isochores from the new table are shown (2.7 and 2.4 g/cm$^3$) that should, if the table is accurate, 
bracket the experimental data.  Note that other experimental data from refs. \cite{desai84,ho83, simmons60} are in
close agreement with the Brandt and Neuer data.
}
\label{fig_desai}
\end{figure}

In the solid regime (figure \ref{fig_models}) we have used the same model as given in reference \cite{lee84}.
The adjustable parameter in that model is tuned to match the data given in reference  \cite{desai84}.  The
melt line is tuned to fit the experiment of reference \cite{leitner17}.  Note that no special consideration 
is given to the solidus-liquidus region, or to the liquid-vapour coexistence region.  The model of reference
\cite{lee84} is simply used below $T_{melt}$, and above $T_{melt}$ the DFT based model is used.  In figure
\ref{fig_desai} we show a comparison of the new table to the experiments of reference \cite{leitner17, brandt07}.
To avoid interpolation artifacts we have shown the two isochores that should bracket the data (which account
for thermal expansion).  Recall that the
solid model was tuned to match the Desai data at melt \cite{desai84}, and the Desai data are in agreement with 
the Brandt data \cite{brandt07}.  The new table does bracket the data for T $>$ 500 K, but below this there
is a small disagreement.  The fault likely lies with the relatively crude solid model that we have used.
Note that Leitner et al comment in their paper \cite{leitner17} that their data in the solid phase should be
regarded as an estimate.  The liquid phase data are thought to be accurate (figure \ref{fig_gathers}).

\section{Discussion\label{sec_dis}}
The new table recovers the known limit at high temperature and agrees reasonable well with the limited 
experimental data available.  Though it interpolates (using Akima interpolation \cite{akima70}) through the
low temperature expanded metal regime, this interpolation was guided by DFT-MD calculations and as a result
is reasonable in this regime.  The table is based on DFT calculations that have no adjustable parameters.
However, the choice of exchange and correlation potential is clearly important in the low temperature
regime (at high temperatures the results are insensitive to this choice).  It seems likely that our DFT-MD
results would be made more accurate by using the HSE functional, though this is much more computationally
expensive and we are already pushing the limits of what is practical for table production.

The simple model that we have used for the solid phase could be improved but such an effort
would only be warranted if lower temperatures (i.e. below 10$^{-2}$ eV $\approx$ 116 K) or
higher accuracy are needed.  Further, we have given no special consideration to the solidus-liquidus region, or
to the coexistence region.  This could be improved, but is difficult in practice due to the finite size
limitations of the DFT approach.

\section{Conclusions\label{sec_con}}
A new Sesame-type table for the electrical conductivity of aluminum has been presented.  The table is
constructed from density-functional theory based calculations.  Comparisons to experiments in the solid, liquid
and plasma regimes show good agreement.  The high temperature, classical limit is recovered.  Avenues for
improvement have been discussed.  The table should be useful for users in the dense plasma community.

\section*{Acknowledgments}
 Los Alamos National Laboratory is operated by Triad National Security, LLC, for the National Nuclear Security Administration of U.S. Department of Energy (Contract No. 89233218NCA000001).
We wish to acknowledge useful conversations with Dr. Suxing Hu.

\appendix
\section{DFT-MD results}
The results for the electrical conductivity from our Kubo-Greenwood, DFT-MD simulations are given in table \ref{tab_cg}.
\setlength{\tabcolsep}{0.5em} 
{\renewcommand{\arraystretch}{2.2}
\begin{table*}[]
\caption{DFT-MD results is units of 10$^6$ S m$^{-1}$.  
The value is obtained from averaging the conductivity over 10 snapshots, and the error 
corresponds to the standard deviation over the 10 measurements. Cases with low density 
and/or high temperature require up to several thousands of bands for the MD alone, and 
were excluded due to the high computational cost. Cases with low temperature and low 
density resulted in solid-like configurations, and were also excluded.
\label{tab_cg}}
\begin{center}
\begin{tabular}{c c | c c c c c c }
&  & \multicolumn{6}{c} {Density [g/cm$^3$]} \\
&  & 0.539  &  0.854  &  1.353  &  1.704  &  2.145  &  2.700 \\
\hline
\multirow{3}{*}{\rotatebox[origin=r]{90}{Temperature [eV]}} & 10.0  & & & & & & 0.844 $\pm$ 0.025  \\
& 2.51  & & 0.308 $\pm$ 0.008 & 0.625 $\pm$ 0.027 & 0.929 $\pm$ 0.066 & 1.319 $\pm$ 0.086 & 1.803 $\pm$ 0.057  \\
& 0.631  & 0.170 $\pm$ 0.003  & 0.342 $\pm$ 0.009 & 0.809 $\pm$ 0.051 & 1.307 $\pm$ 0.091 & 1.936 $\pm$ 0.167 & 2.817 $\pm$ 0.170  \\
& 0.251  &                    &                   &                   & 1.654 $\pm$ 0.101 & 2.759 $\pm$ 0.155 & 3.45 $\pm$ 0.1 
\end{tabular}
\end{center}
\end{table*}

\bibliographystyle{unsrt}
\bibliography{phys_bib}

\end{document}